# ANTI-SYNCHRONIZATION IN SOME MULTIPLE TIME DELAY POWER SYSTEMS

E.M.Shahverdiev[1]

Institute of Physics,33,H.Javid Avenue,Baku,AZ1143, Azerbaijan

## ABSTRACT

We investigate chaos antisynchronization between two uni-directionally coupled multiple time delay power systems.The results are of certain importance to prevent power black-out in the entire power grid.

## I.INTRODUCTION

Chaos synchronization [1] is one of the basic features in nonlinear science and it is one of fundamental importance in a variety of complex physical, chemical,power and biological systems with possible application areas in secure communications, optimization of nonlinear system performance, modeling brain activity and pattern recognition phenomena,avoiding power black-out [1].

Finite signal transmission times, switching speeds and memory effects makes time-delayed systems (DDE) ubiquitous in nature, technology and society [2]. Therefore the study of synchronization phenomena in such systems is of considerable practical importance [1].

Electrical power systems are essentially nonlinear dynamical systems [3]. Study of chaos and its control in such systems [3-4] could be of considerable importance from the point of view of avoiding undesirable behaviours such as power black-out. Synchronization in power systems is of huge importance from the management point of view of complex power systems. This paper is devoted to chaos anti-synchronization in some simple power models- a single -machine-infinite-bus systems (SMIB) power system [3-6]. We study chaos anti-synchronization in power systems described by both linearly and non-linearly coupled multiple time delay power systems.

## II.CHAOS ANTI-SYNCHRONIZATION IN MULTIPLE TIME DELAY POWER SYSTEMS

Consider the classical SMIB power system [3-6], called swing equation generalised for two time delays

$$M\frac{d\theta^2}{dt^2}+D\frac{d\theta}{dt}+P_{max}\sin\theta = P_m+P_{\tau_1}+P_{\tau_2},\qquad(1)$$

---

[1]e-mail:shahverdiev@physics.ab.az

which can be written as a system of first order equations:

$$\frac{dx_1}{dt} = x_2, \tag{2}$$

$$\frac{dx_2}{dt} = -cx_2 - \beta \sin x_1 + f \sin \omega t + k_1 \sin(R_1 x_1(t-\tau_1)) + k_2 \sin(R_2 x_1(t-\tau_2)), \tag{3}$$

where

$$x_1 = \theta, x_2 = \frac{d\theta}{dt}, c = \frac{D}{M}, \beta = \frac{P_{max}}{M}, f = \frac{A}{M}, \tag{4}$$

Here $M$ is the moment of inertia,$D$ is the damping constant,$P_m$ is the power of the machine,$P_{max}$ is the maximum power of generator, and $P_m = A \sin \omega t$. $P_{\tau_1} = m_1 \sin(R_1 x_1(t-\tau_1)), P_{\tau_2} = m_2 \sin(R_2 x_1(t-\tau_2))$ are the delayed fedback terms with $k_1 = m_1/M$ and $k_2 = m_2/M$.

Dot means a differentiation with respect to time. Introducing a new state variable via $x_3 = \omega t$ this SMIB power system can be written as a system of three differential equations without variable time term in the right-hand side:

$$\frac{dx_1}{dt} = x_2, \tag{5}$$

$$\frac{dx_2}{dt} = -cx_2 - \beta \sin x_1 + f \sin x_3 + k_1 \sin(R_1 x_1(t-\tau_1)) + k_2 \sin(R_2 x_1(t-\tau_2)), \tag{6}$$

$$\frac{dx_3}{dt} = \omega, \tag{7}$$

Extensive investigation of dynamical behaviour of this power system is conducted in [7]. In this paper our first task is to investigate chaos anti-synchronization between linearly coupled multiple time delay power systems, e.g. $x_2 = -y_2$. For that purpose consider the following DDE system coupled with the system of Eqs.(5-7) by the coupling term $K(x_2 - y_2)$

$$\frac{dy_1}{dt} = y_2, \tag{8}$$

$$\frac{dy_2}{dt} = -cy_2 - \beta \sin y_1 + f \sin y_3 + k_3 \sin(R_1 y_1(t-\tau_1)) + k_4 \sin(R_2 y_1(t-\tau_2)) + K(x_2 - y_2), \tag{9}$$

$$\frac{dy_3}{dt} = \omega, \tag{10}$$

Existence condition for the anti-synchronization regime can be found as in references [2]:$k_1 = k_3, k_2 = k_4$. Then we will consider the case of nonlinearly coupled power systems. For this case Eq.(9) can be written as

$$\frac{dy_2}{dt} = -cy_2 - \beta \sin y_1 + f \sin y_3 + k_3 \sin(R_1 y_1(t-\tau_1)) + k_4 \sin(R_2 y_1(t-\tau_2)) + k_5 \sin(R_1 x_1(t-\tau_3)), \tag{11}$$

Here $\tau_3$ is the coupling delay time between the power systems;$k_5$ is the coupling strength between the systems. Existence conditions for anti-synchronization can be estyablished as ($k_1 = k_3 + k_5$ and $k_2 = k_4$ or $k_1 = k_3$ and $k_2 = k_4 + k_5$). We note that in case of nonlinearly multiple time delay systems the number of poosible synchronization regimes can be inreased significantly, see, e.g. references in [2].

## III.NUMERICAL SIMULATIONS

Now we demonstrate that two multiple time delay coupled power systems can be anti-synchronized. We use DDE software built-in in Matlab 7 for the numerical simulations.We study the synchronization between the power systems using the correlation coefficient $C$ [8]

$$C(\Delta t) = \frac{<(x(t)-<x>)(y(t+\Delta t)-<y>)>}{\sqrt{<(x(t)-<x>)^2><(y(t+\Delta t)-<y>)^2>}}$$

where $x$ and $y$ are the outputs of the interacting systems; the brackets$<.>$ represent the time average; $\Delta t$ is a time shift between the system outputs:for our case$\Delta t = 0$. This coefficient indicates the quality of synchronization:$C = -1$ means perfect anti-synchronization.

Fig.1 demonstrates the dynamics of variable $x_2$ for Eqs.(5-7) and $C$ shows the quality of anti-synchronization between the linearly coupled multiple time delay power systems $x_2 = -y_2$, Eqs.(5-7) and Eqs.(8-10)($x_2 = -y_2$) . Figure 2 depicts the dynamics of variable $x_2$ for system of Eqs.(5-7) with $C$ indicating the level of anti-synchronization between the nonlinearly coupled power systems, Eqs(5-7) and Eqs.(8,10,11). The system parameters are: with existence condition of the antisynchronization regime $k_1 = k_3 + k_5$ and $k_2 = k_4$. In figure 3 we present example of anti-synchronization with existence conditions $k_1 = k_3$ and $k_2 = k_4 + k_5$, the other paprameters as in figure 2.

We also note that the amplitude of the change of the dynamical variables, e.g.Eqs(5-7) for the power systems can be governed by the paprameter changes in the system. In figures 4-7 we present several scenarios for such changes. Figure 4(a) describes dynamics of $x_2$ for the damping constant $c = 2$ and $\beta = 1, m = 1, \omega = 1, R_1 = R_2 = 5, k_1 == 0.2, k_2 = 15, \tau_1 = 0.1, \tau_2 = 0.2$. Fig.4(b) corresponds to the case of $c = 4$, with $c = 6$ for fig.4(c).The other parameters as in figure 4(a).
In figures 5(a)-5(c) we study the effect of the maximum power generator on the dynamics of $x_2$.Fig.5(a), 5(b) and 5(c) corresponds to b=2, b=4 and b=6, respectively; the other parameters as for figures 4.
Figures 6(a-c) show the role of the feedback strength in the dynamics of the power system, Eqs.(5-7);for fig.6(a) $k_2 = 1$, for fig.6(b) $k_2 = 5$, and for fig.6(c)-$k_2 = 15$;the other parameters as in fogures 4.
In figures 7(a-c) the influence of the frequency modulation of the feedback $G$ on the power systems dynamics is depicted;$G = 0.0005$ for fig.7(a), $G = 0.5$ for fig.7(b) and $G = 15$ for fig.7(c) with other paprameters as in for figures 4. These results reveal powerful of of the paprameter changes in the dynamics of the power systems. Detailed study of this question will be reported elsewhere.

## IV. DISCUSSION AND CONCLUSIONS

The results of the investigation carried out in this paper underscore the principal possibility of anti-synchronization between the power systems coupled both linearly and nonlinearly. This testifies to the wide range applicability of the results in the practical situations. The main conclusion of the paper is the possibility of avoiding possible power black-out in the power grid by anti-synchronising different parts of the grid. Although we have presented the results for the chaotic behaviour, the conclusion is equally applicable to the case of non-chaotic systems. Also,as established above the parameter changes can tame unruly amplitude swings, and thus can be used for the instability control in the power grid.

## V. ACKNOWLEDGEMENTS

This research was supported by a Marie Curie Action within the $6^{th}$ European Community Framework Programme Contract N:MIF2-CT-2007-039927-980065.

FIGURE CAPTIONS

FIG.1.Numerical simulation of the linearly coupled multiple time delay power models,Eqs.(5-7) and Eqs.(8-10):time series of $x_2(t)$ is shown. $C$ is the correlation coefficient between the power systems:$x_2 = -y_2$. The parameters are $c = 65, \beta = 1, m = 1, \omega = 1, R_1 = R_2 = 0.005, k_1 = k_3 = 0.2, k_2 = k_4 = 0.4, \tau_1 = 0.6, \tau_2 = 0.2, K = 100$.

FIG.2.Numerical simulation of the non-linearly coupled multiple time delay power models,Eqs.(5-7) and Eqs.(8,10,11):time series of $x_2(t)$ is shown. $C$ is the correlation coefficient between the power systems:$x_2 = -y_2$ The parameters are $c = 1, \beta = 1, m = 1, \omega = 1, R_1 = R_2 = 5, k_1 = 20, k_3 = 0.1, k_5 = 19.9, k_2 = k_4 = 0.4, \tau_1 = 0.1, \tau_2 = 0.2, \tau_3 = 0.3$.

FIG.3.Numerical simulation of the non-linearly coupled multiple time delay power models,Eqs.(5-7) and Eqs.(8,10,11):time series of $x_2(t)$ is shown. $C$ is the correlation coefficient between the power systems:$x_2 = -y_2$ The parameters are $c = 1, \beta = 1, m = 1, \omega = 1, R_1 = R_2 = 5, k_1 = k_3 = 0.2, k_2 = 15, k_4 = 0.1, k_5 = 14.9, \tau_1 = 0.1, \tau_2 = 0.2, \tau_3 = 0.3$.

FIG.4.Numerical simulation of the DDE power model,Eqs.(5-7):time series of $x_2(t)$ are shown for the several parameter changes:c=2 for fig.4(a);c=4 for fig.4(b); c=6 for fig.4(c). The other parameters are:$\beta = 1, m = 1, \omega = 1, R_1 = R_2 = 5, k_1 = 0.2, k_2 = 15, \tau_1 = 0.1, \tau_2 = 0.2$.

FIG.5.Numerical simulation of the DDE power model,Eqs.(5-7):time series of $x_2(t)$ are shown for the several parameter changes:b=2 for fig.5(a);b=4 for fig.5(b); b=8 for fig.5(c). The other parameters as in for figures 4.

FIG.6.Numerical simulation of the DDE power model,Eqs.(5-7):time series of $x_2(t)$ are shown for the several parameter changes:$k_2 = 1$ for fig.6(a);$k_2 = 5$ for fig.6(b); $k_2 = 15$ for fig.6(c). The other parameters as in for figures 4.

FIG.7.Numerical simulation of the DDE power model,Eqs.(5-7):time series of $x_2(t)$ are shown for the several parameter changes:$G = 0.0005$ for fig.7(a);$G = 0.5$ for fig.7(b); $G = 15$ for fig.7(c). The other parameters as in for figures 4.

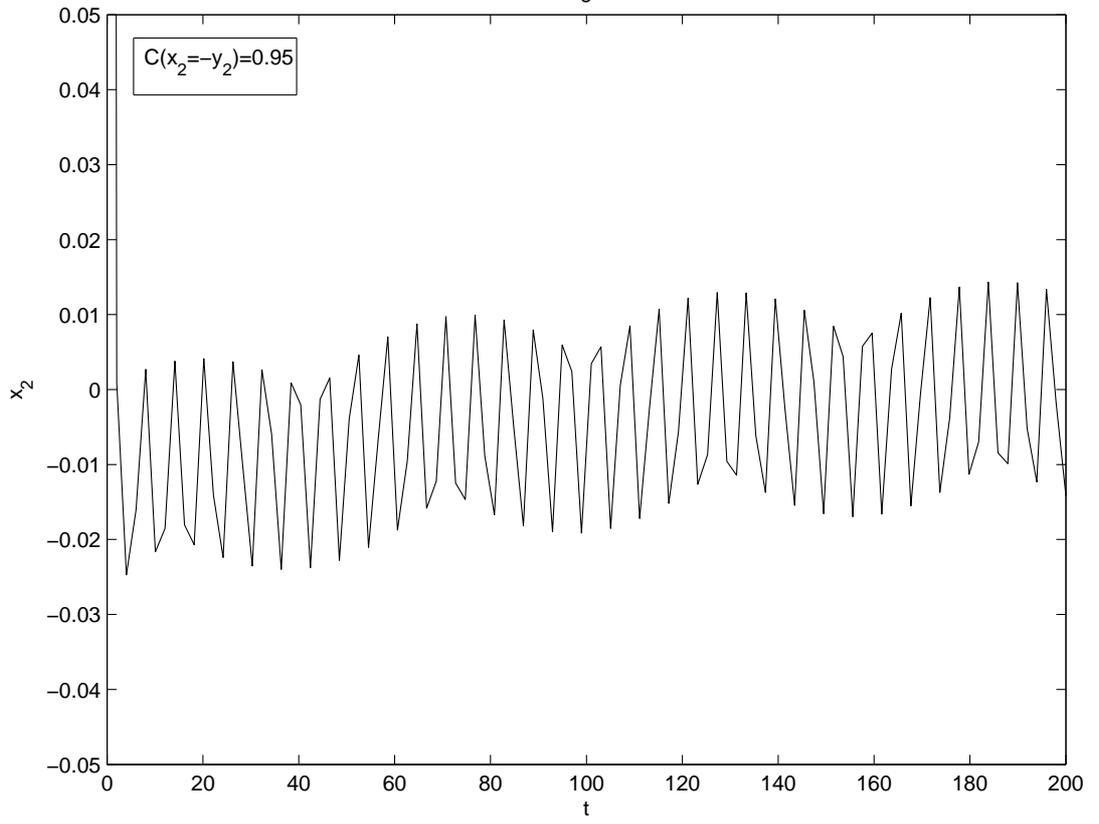

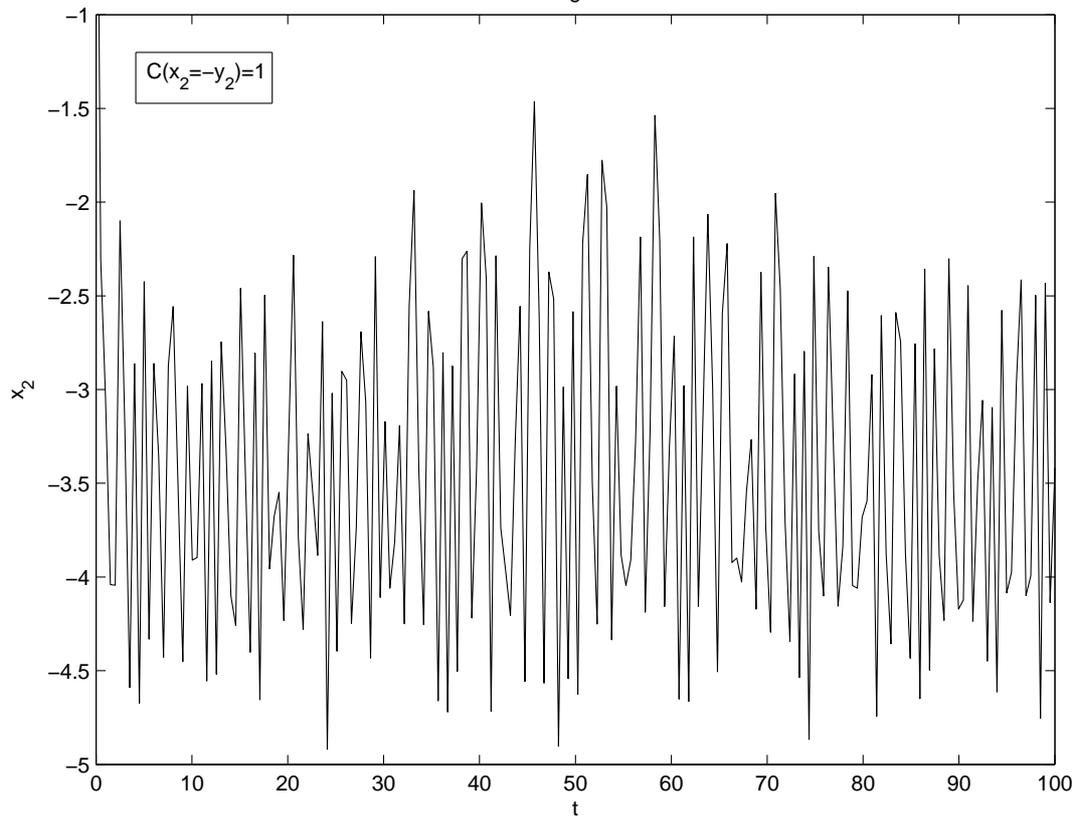

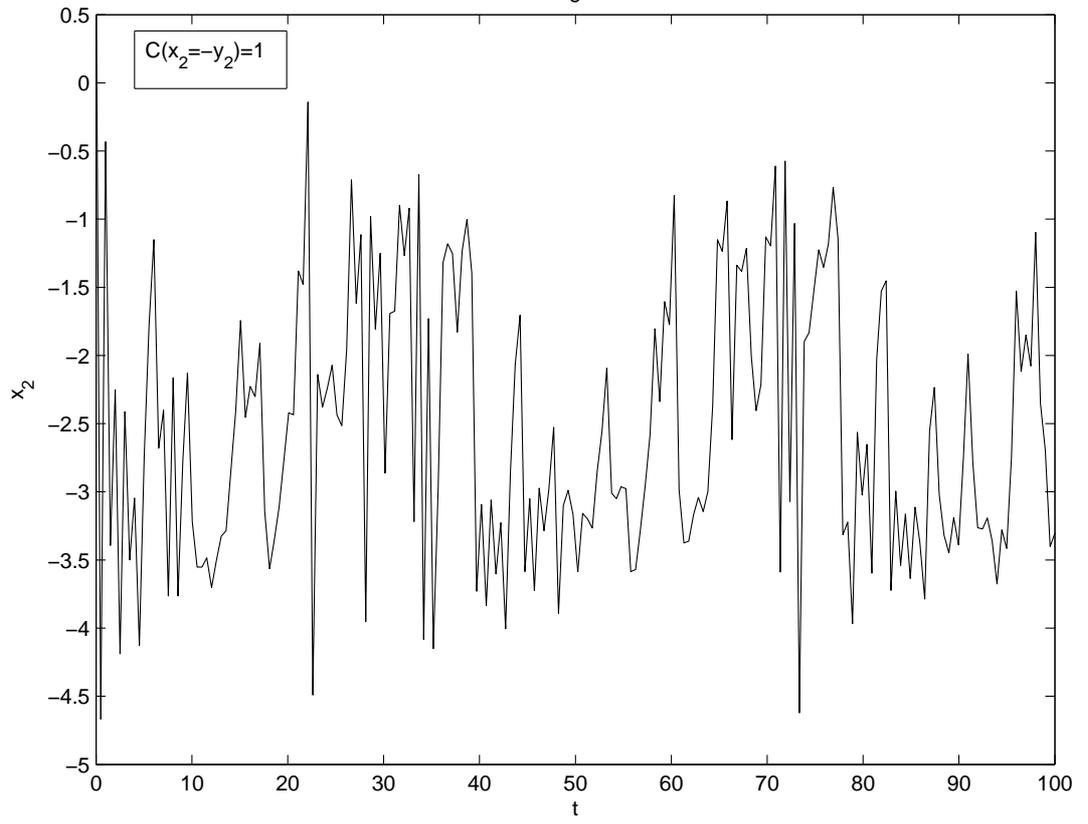

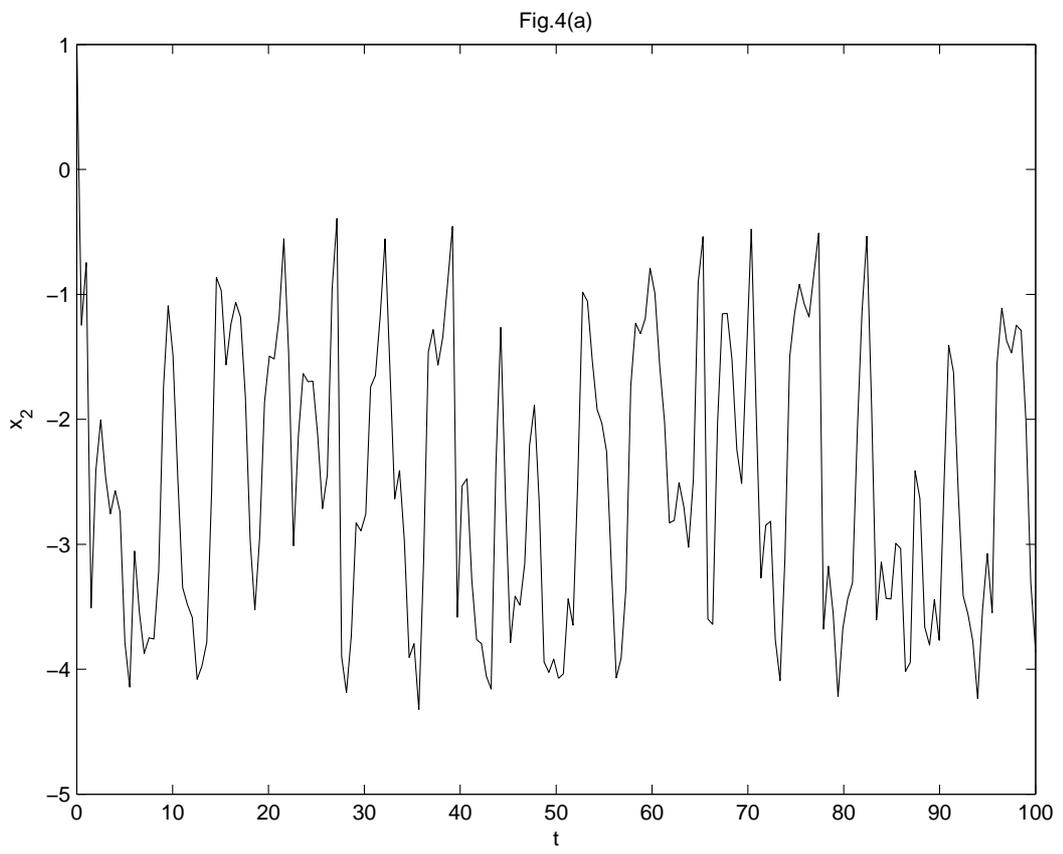

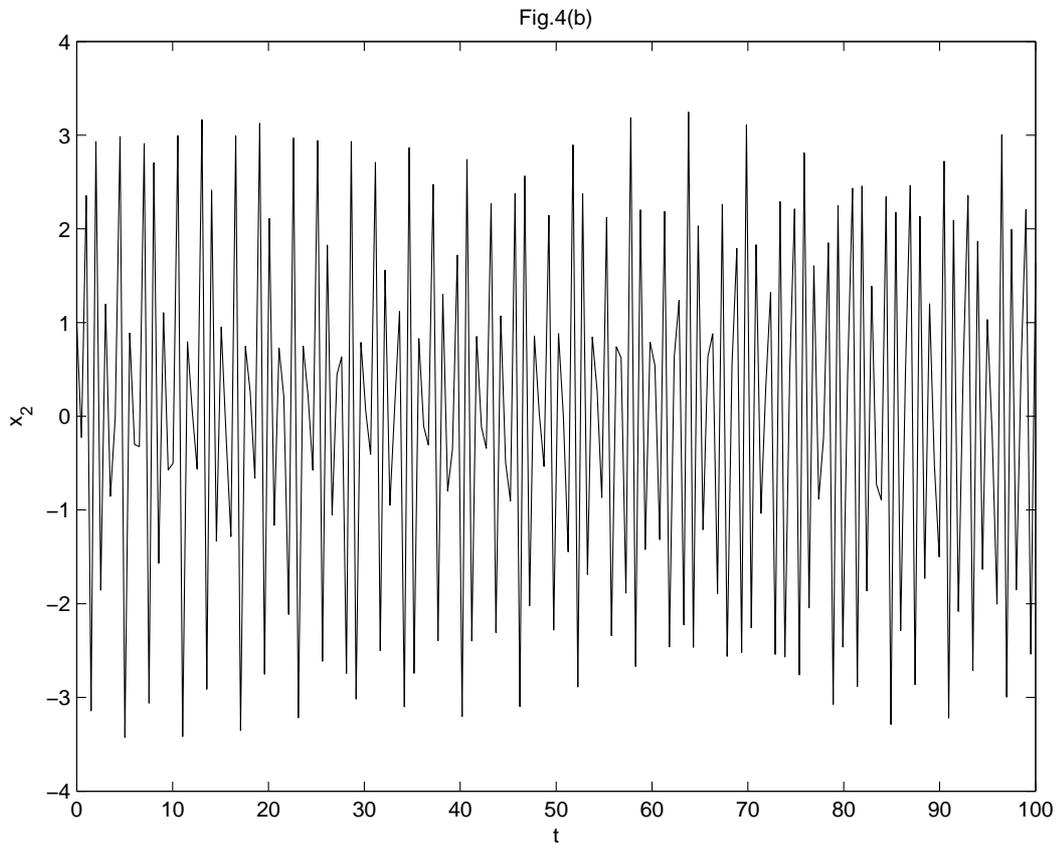

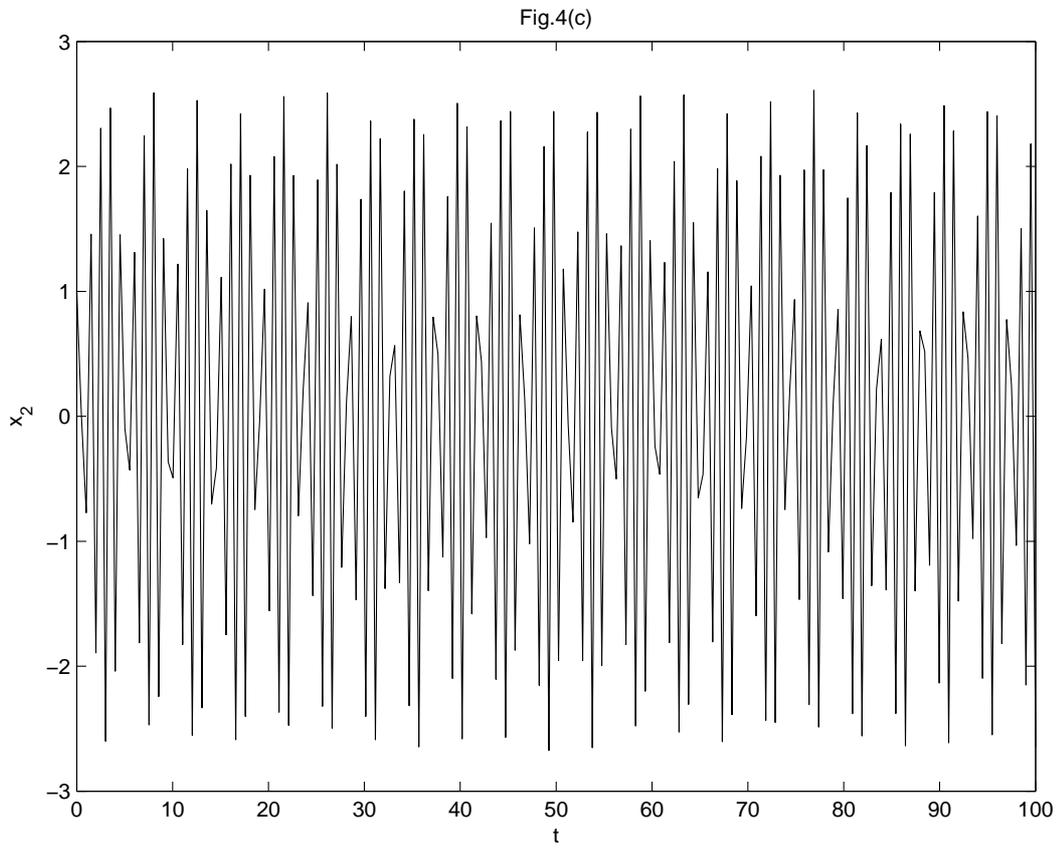

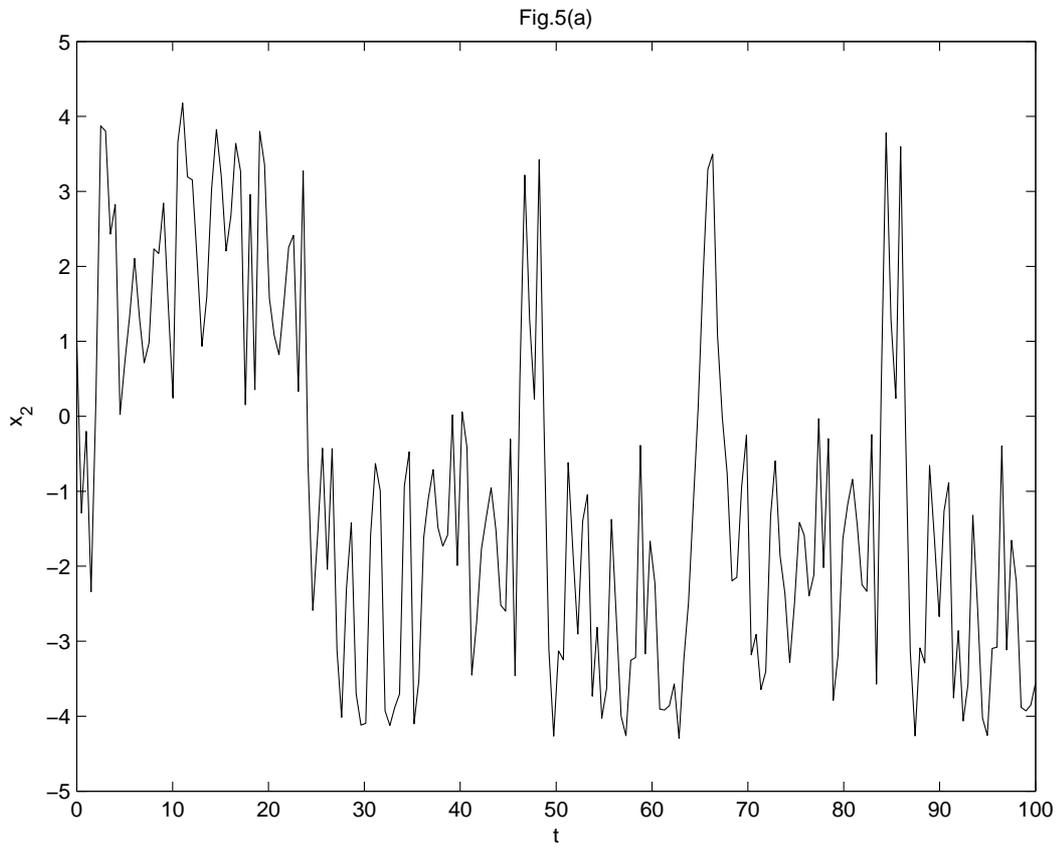

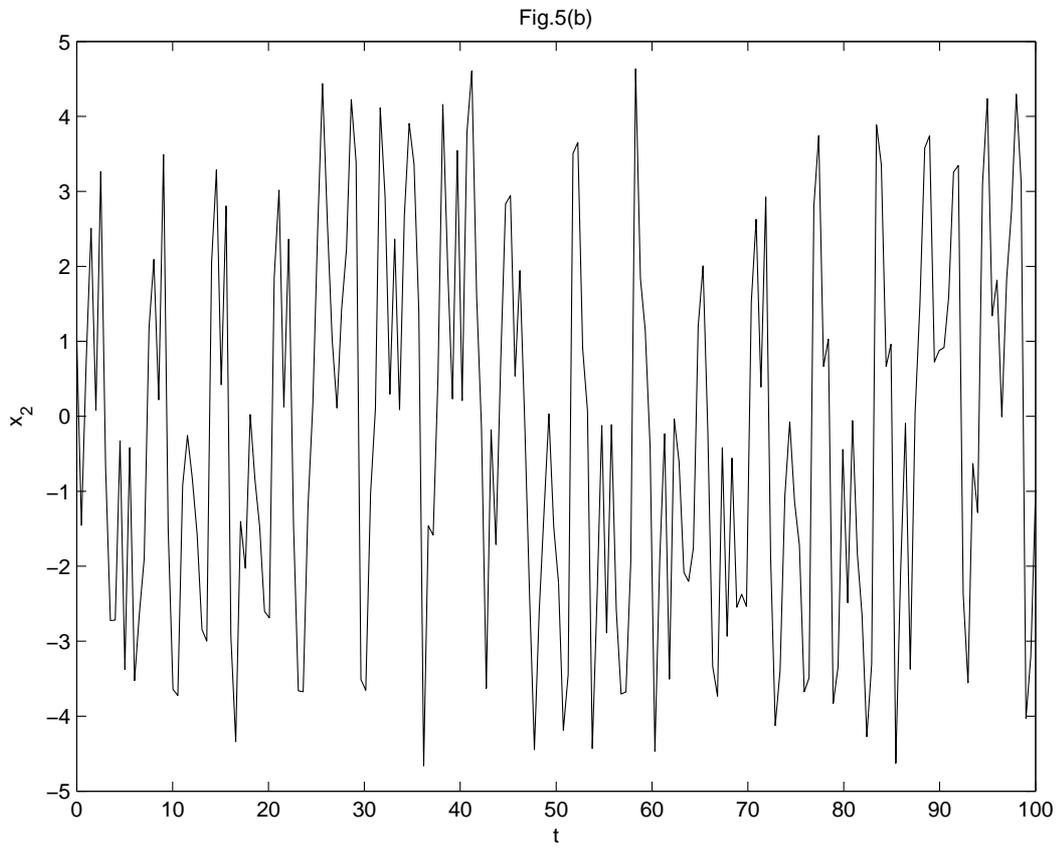

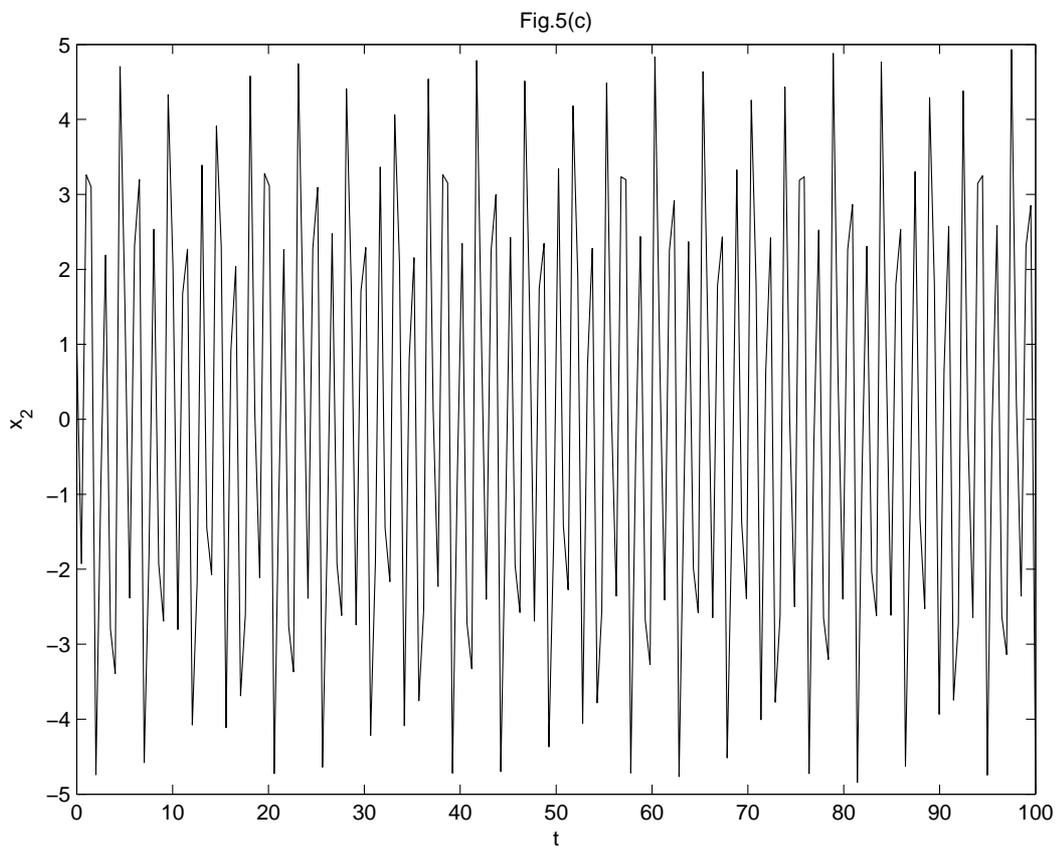

Fig.5(c)

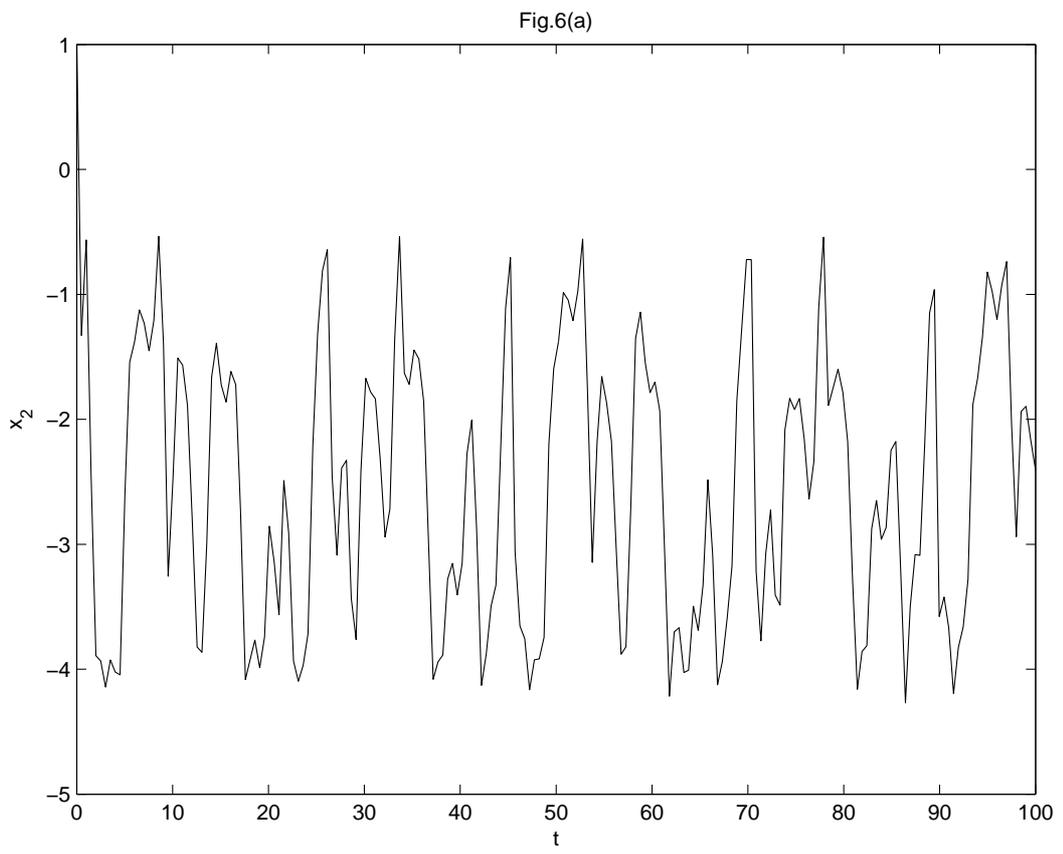

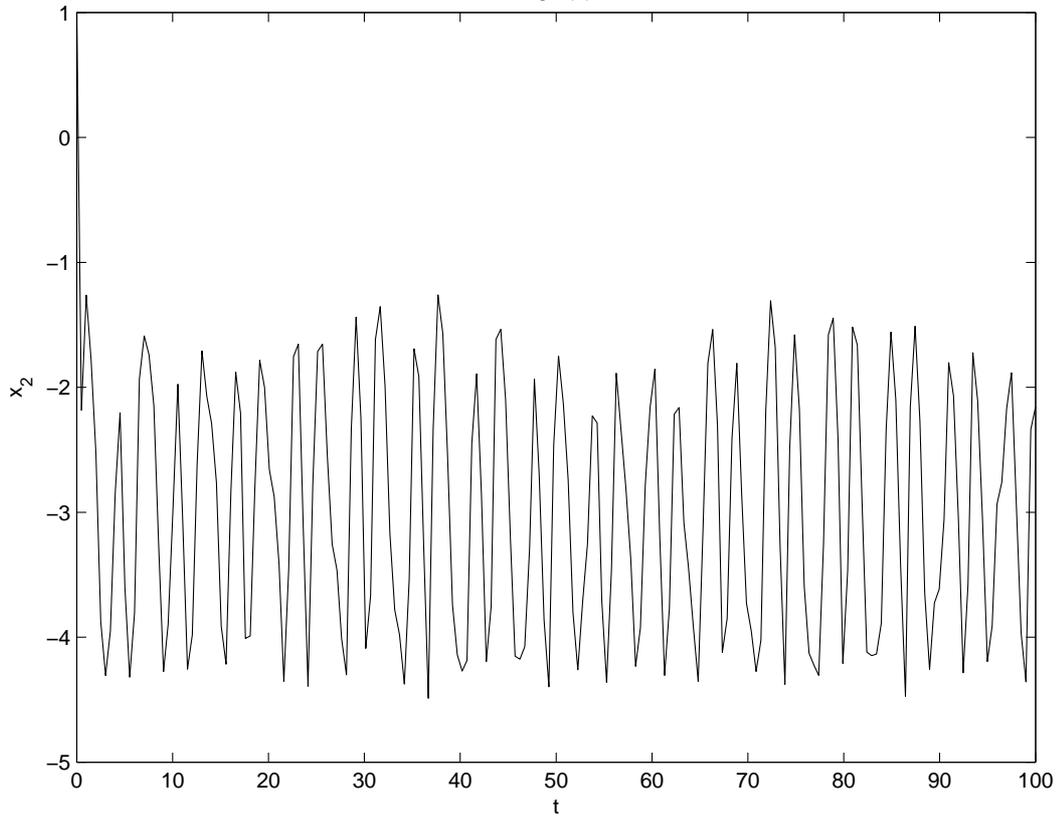

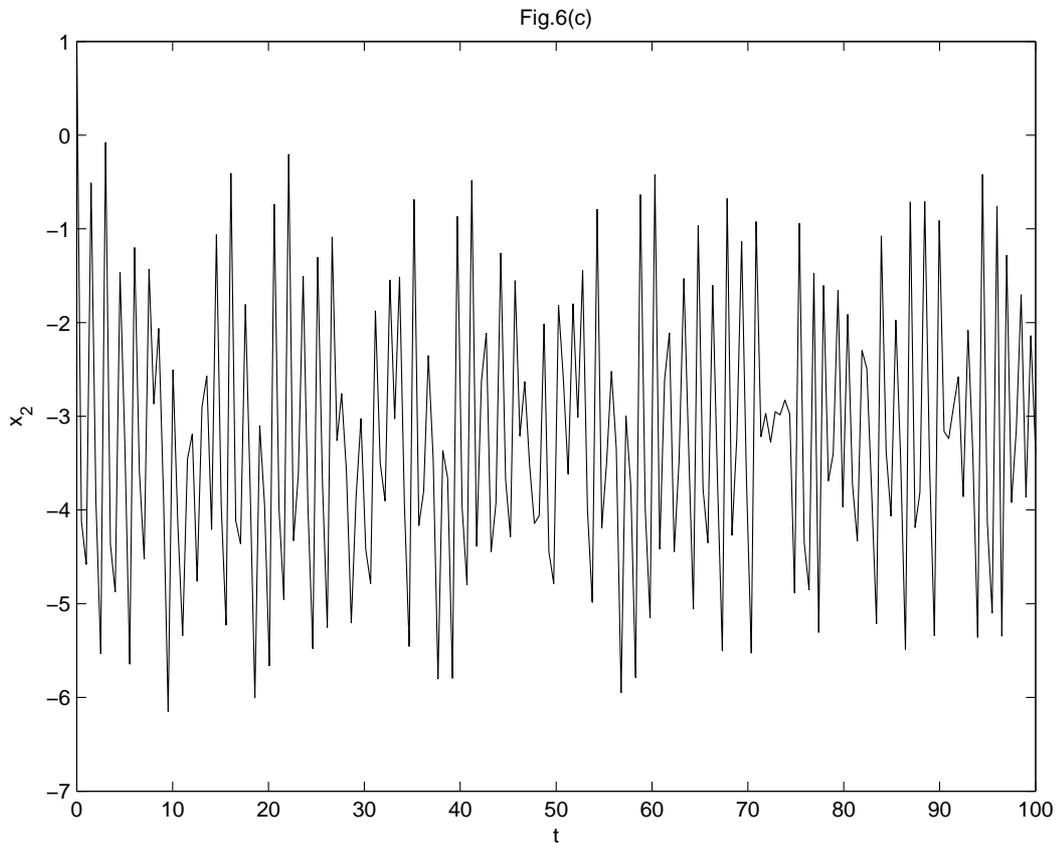

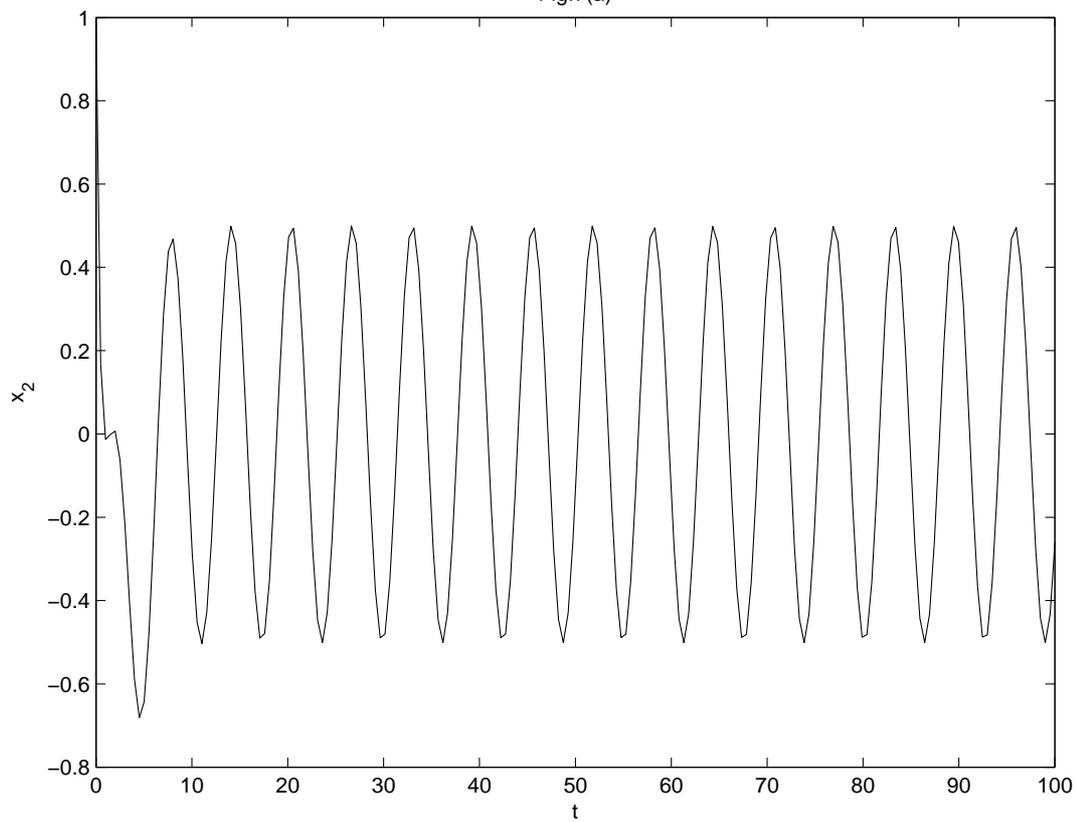

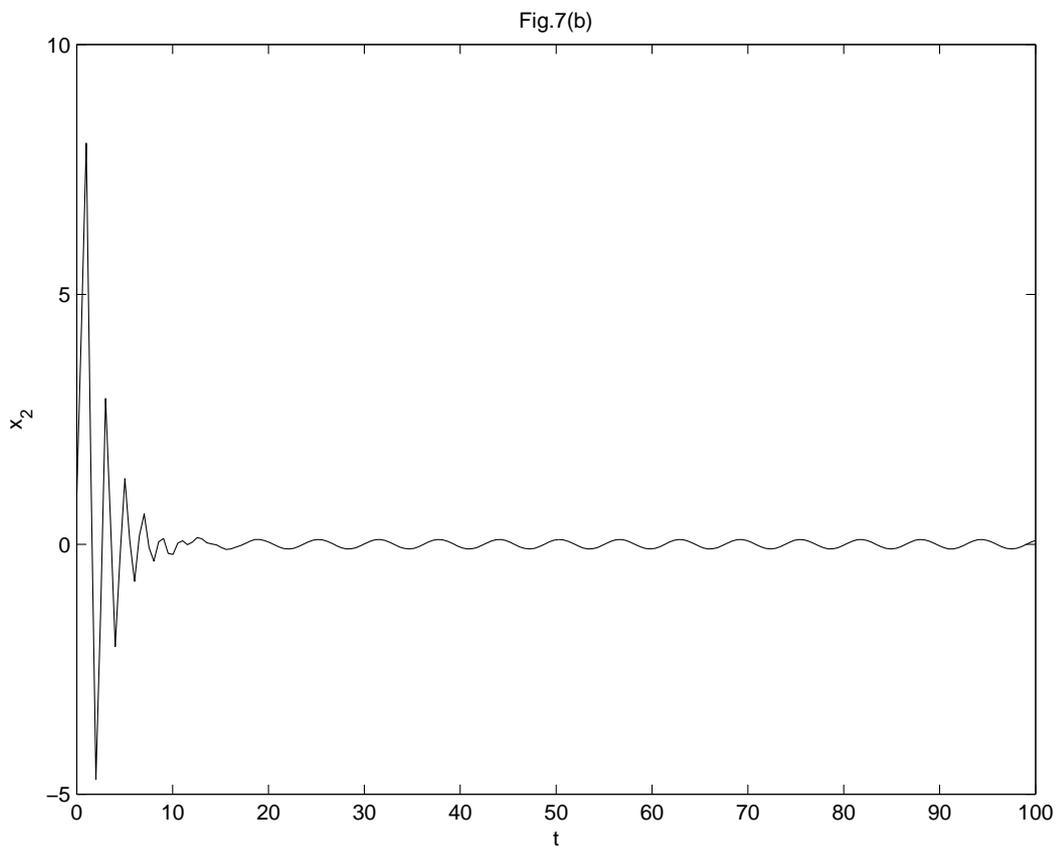

Fig.7(b)

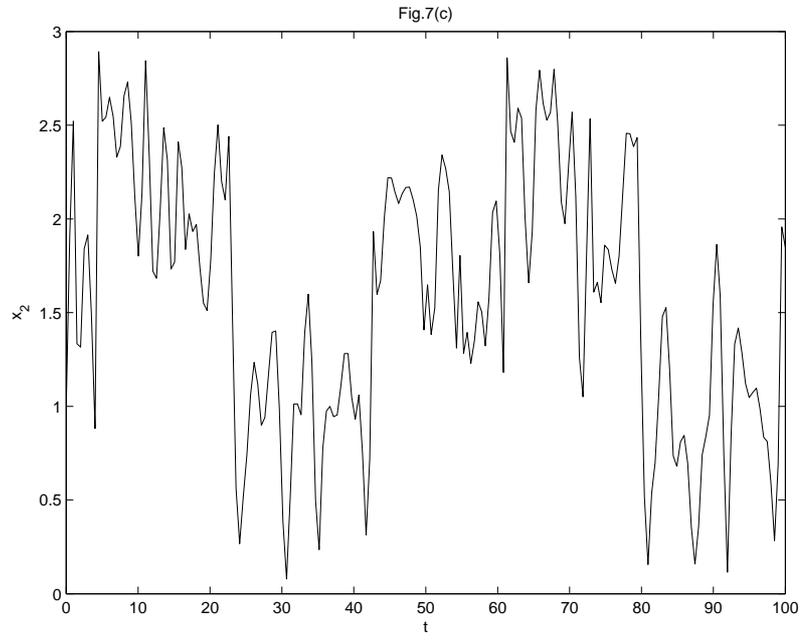

Fig.7(c)